\documentstyle[twocolumn,epsfig,prl,aps]{revtex}
\begin{document}
\draft
\newcommand{\kpme}{\emph{$K_{\pi\mu e}$}}
\newcommand{\kpkpee}{\emph{$K_{\pi e e}$}}
\newcommand{\kppp}{\emph{$K_{\tau}$}}
\newcommand{\kpp}{$K^{+}\rightarrow\pi^{+}\pi^{0}$}
\newcommand{\peeg}{$\pi^{0}\rightarrow e^{+}e^{-}\gamma$}
\newcommand{\penu}{$\pi^{-}\rightarrow e^{-}\overline{\nu}$}
\newcommand{\pmunu}{$\pi^{+}\rightarrow \mu^{+}\nu$}
\newcommand{\kdal}{\emph{$K_{Dal}$}}
\newcommand{\kpee}{$K^{+}\rightarrow\pi^{+}e^+e^-$}
\newcommand{\kpimumu}{$K^{+}\rightarrow\pi^{+}\mu^+\mu^-$}
\newcommand{\ktau}{$K^{+}\rightarrow\pi^{+}\pi^+\pi^-$}
\newcommand{\kpimue}{$K^{+}\rightarrow\pi^{+}\mu^+e^-$}
\newcommand{\dal}{$K_{dal}$}
\newcommand{\pee}{$K_{\pi e e}$}
\newcommand{\taus}{$K_{\tau}$}

\wideabs{
\title{
A new measurement of the properties of the rare decay 
$K^+\rightarrow \pi^+ e^+ e^-$
}
\author{
R. Appel$^{6,3}$, G.S. Atoyan$^4$, B. Bassalleck$^2$,  
D.R. Bergman$^6$\cite{DB}, D.N. Brown$^3$\cite{DNB}, N. Cheung$^3$, 
S. Dhawan$^6$,  \\
H. Do$^6$, J. Egger$^5$, S. Eilerts$^2$\cite{SE}, C. Felder$^{3,1}$,   
H. Fischer$^2$\cite{HF}, M. Gach$^3$\cite{MG}, W. Herold$^5$, \\  
V.V. Issakov$^4$, H. Kaspar$^5$, D.E. Kraus$^3$, D. M. Lazarus$^1$, 
L. Leipuner$^1$, P. Lichard$^3$, J. Lowe$^2$, J. Lozano$^6$\cite{JL},\\  
 H. Ma$^1$, W. Majid$^6$\cite{WMa}, W. Menzel$^7$\cite{WMe}, 
S. Pislak$^{8,6}$, A.A. Poblaguev$^4$,   
V.E. Postoev$^4$, A.L. Proskurjakov$^4$, \\ P. Rehak$^1$,  
P. Robmann$^8$, A. Sher$^3$, T.L. Thomas$^2$, J.A. Thompson$^3$, 
P. Tru\"ol$^{8,6}$, H. Weyer$^{7,5}$, and M.E. Zeller$^6$   \\
}

\address{
$^1$ Brookhaven National Laboratory, Upton L. I., NY 11973, USA\\ 
$^2$Department of Physics and Astronomy, 
University of New Mexico, Albuquerque, NM 87131, USA\\
$^3$ Department of Physics and Astronomy, University of Pittsburgh,
Pittsburgh, PA 15260, USA \\ 
$^4$Institute for Nuclear Research of Russian Academy of Sciences, 
Moscow 117 312, Russia \\
$^5$Paul Scherrer Institut, CH-5232 Villigen, Switzerland\\ 
$^6$ Physics Department, Yale University, New Haven, CT 06511, USA\\
$^7$Physikalisches Institut, Universit\"at Basel, CH-4046 Basel,Switzerland\\
$^8$ Physik-Institut, Universit\"at Z\"urich, CH-8057 Z\"urich, Switzerland}
\date{\today}
\maketitle

\begin{abstract}
A large low-background sample of events (10300) has been
collected for the rare decay of kaons in flight  
$K^+\rightarrow \pi^+e^+e^-$ by experiment E865 at the Brookhaven AGS. 
The decay products were accepted by a
broad band high-resolution charged particle spectrometer with particle
identification.
The branching ratio 
$(2.94\pm 0.05\,(stat.)\pm 0.13\,(syst.)\pm 0.05\,(model))\times 10^{-7}$ 
was determined normalizing to events from the decay chain
$K^+\rightarrow\pi^+\pi^0;\,\pi^0\rightarrow e^+e^-\gamma$.
From the analysis of the decay distributions the vector nature
of this decay is firmly established now, and limits on scalar and
tensor contributions are deduced. From the $e^+e^-$ invariant mass
distribution the decay form factor $f(z)=f_0(1+\delta z)$ 
($z=M_{ee}^2/m_K^2$) is determined with 
$\delta=2.14\pm 0.13\pm 0.15$.
Chiral QCD perturbation theory predictions for the 
form factor are also tested, and
terms beyond leading order ${\cal O}(p^4)$ are found to be important.
\end{abstract}

}

The decay $K^+\rightarrow \pi^+ e^+ e^-$ ($K_{\pi e e}$)
proceeds via a flavor changing neutral current and is highly suppressed
by the GIM-mechanism.  The decay
rate was first calculated \cite{gaillard} 
assuming a short-distance
$s\rightarrow d \gamma$ transition
at the quark level.  Later it was 
realized that the long-distance effects dominate the decay mechanism
\cite{theory}. Several recent calculations,
which study the rate and invariant 
electron-positron mass ($M_{ee}$) distributions, 
were performed within the framework of chiral QCD perturbation theory (ChPT)
\cite{ecker,donoghue,dambrosio},
an approach which has been quite successful at modeling many decay modes of 
the light mesons \cite{chpt}. 

The first few 
events for this decay were observed at CERN~\cite{bloch}.
Two subsequent experiments at the Brook\-haven AGS, E777~\cite{alliegro} and 
E851~\cite{deshpande}, observed 500 and 800 events, and measured  
branching ratios of $2.75\pm 0.26$ and
$2.81\pm 0.20$ ($\times 10^{-7}$), respectively. We report here the 
results of a new measurement with larger acceptance and 
significantly increased statistics. 
Our data
allow a detailed study of the decay form factor and a
comparison with ChPT calculations and other models.

\begin{figure}[htb]
\epsfig{figure=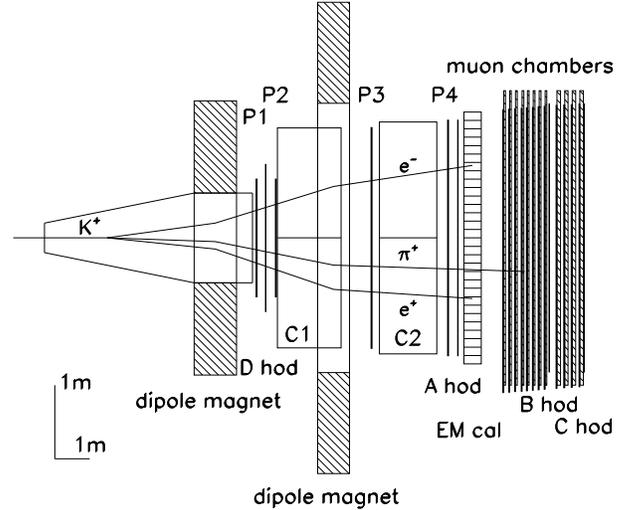,width=80mm}\centering
\caption[Plan view of the E865 detector]
{Plan view of the E865 detector. P1 - P4: proportional chambers;
C1, C2: \v{C}erenkov counters; ABCD: scintillator
hodoscopes. The beam passes through holes in the calorimeter
and muon stack, dead regions in the proportional chambers, 
He bags and H$_2$ filled beam tubes in the 
\v{C}erenkov counters, which are not shown here.} 
\label{fig:detector}
\end{figure}

The experiment was performed at the Brook\-haven National Laboratory's AGS.  
The apparatus, a schematic drawing of which is 
shown in Fig. \ref{fig:detector}, 
was constructed to search for the decay K$^+\rightarrow \pi^+ \mu^+ e^-$ 
\cite{e865,pislak,bergman,eilerts}.  It 
resided in a 6 GeV/c unseparated beam containing 
about $10^8$ K$^+$ and $2\times
10^9$ $\pi^+$ and protons per 1.6 second AGS pulse.  
Downstream of a 5m-long evacuated decay volume 
a dipole magnet separated the trajectories 
by charge, with negative particles going mainly to the left.
This was followed by 
a spectrometer consisting of four proportional chambers 
surrounding a dipole magnet with a 0.833 Tm field integral, which 
determined the momenta and trajectories of
the decay products.  
Particle identification was accomplished with 
\v{C}erenkov
counters filled with H$_2$ on the left (C1L and C2L), and 
CH$_4$ on the right (C1R and C2R) at atmospheric
pressure; with an electromagnetic calorimeter of the Shashlik
design\cite{atoyan}, consisting of of 600 modules, each 11.4 cm by 11.4 cm by 
15 radiation lengths in depth,
arrayed 30 horizontally and 20 vertically;
and with a muon range stack consisting of 24 planes of proportional 
tubes situated 
between iron plates.  With these components, electrons were identified as 
having light in the appropriate \v{C}erenkov counters 
and energy in the calorimeter
consistent with the measured momentum of the
trajectory.  Pions were identified as having no light in 
the \v{C}erenkov counters, and an energy loss in the calorimeter 
consistent with that of a minimum ionizing particle or a hadron
shower.
Compared to the most recent experiment \cite{alliegro,deshpande}
the E865 spectrometer had improved particle identification 
capabilities, spectrometer resolution and larger, more uniform
acceptance.

\begin{figure}[htb]
\epsfig{figure=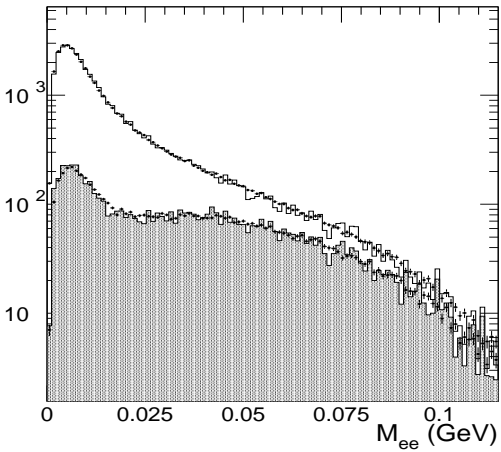,width=80mm,height=70mm}
\centering
\caption[Dalitz decay mass distribution]{
$M_{ee}$ distribution for accepted \dal\ events with 
(lower curve) and without (upper curve) high-mass trigger
required. The Monte Carlo simulation (histogram) includes
events from other $K^+$ decays with a $\pi^0$ in the final state.
}
\label{fig:dalitz}
\end{figure}

The first-level trigger for the experiment is based on 
three charged particle hits using hodoscopes  
D and A (arrays of 10 and 15 slats each on either side of the 
beam line, respectively), and the 
calorimeter.  Most of the rate at this stage comes from
accidentals, and \ktau\ (\taus ) decays. 
For \pee\ decays the next level trigger
required \v{C}erenkov counter signals on each side
of the detector. 
This trigger was dominated by 
events from the decay chain
$K^+\rightarrow \pi^+ \pi^0;\ \pi^0\rightarrow e^+ e^- \gamma$
(\dal ) with low invariant mass $M_{ee}$.
To enhance the \pee\ fraction a high mass trigger
was configured, in which events with
small vertical separation of calorimeter hits were
prescaled. This reduced the number of \dal\ triggers, while keeping
85~\% of the high $M_{ee}$ events. Figure \ref{fig:dalitz}
shows the invariant mass spectra for the prescaled
monitor \dal\ events and those,
which were accepted by this trigger and were used for normalisation.

\begin{figure}[htb]
\epsfig{figure=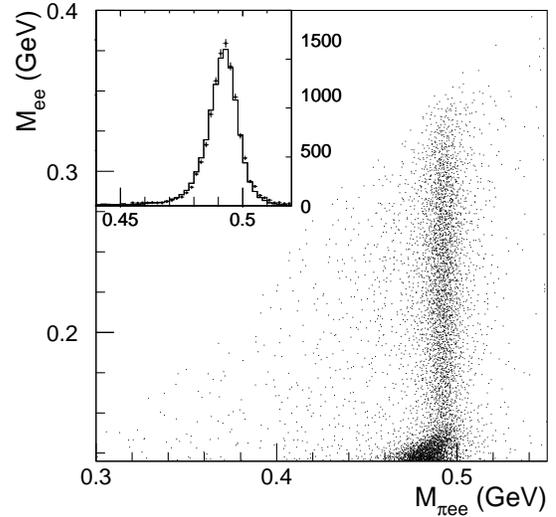,width=80mm}
\centering
\caption[]{Scatter plot $M_{ee}$ versus $M_{\pi  ee}$ 
for \pee\ candidates. Insert: $M_{\pi ee}$ 
mass for \pee\ candidates with $M_{ee}>0.15$ GeV. The
histogram shows the Monte Carlo simulation.}
\label{fig:peems2}
\end{figure}

In the analysis, the basis for the selection of both normalisation 
and signal events was the unambiguous identification of a positive 
pion and lepton pair with trajectories from a common vertex 
located within the decay volume.  
With a cut $M_{ee}>0.15$ GeV we have a nearly pure
sample of \pee\ events with a three particle invariant mass
$M_{\pi ee}\approx m_K$, as shown in Fig. \ref{fig:peems2}. 
The onset of the \dal\ events can be seen
for $M_{ee}<0.13$ GeV and $M_{\pi ee}<m_K$, the latter
because of a missing photon.
For  \pee\ events, the reconstructed $K^+$ was required 
to come from the production target within the limits inferred
from studying \taus\ decays.
Our final signal sample contains 10300 \pee\ candidates including
1.2 \% background events. Our normalisation sample,
after prescaling, contains 10$^5$ \dal\ candidates including a 17 \%
contribution from two other $K^+$ decays giving $\pi^0$-Dalitz pairs 
($K^+\rightarrow\pi^0\mu^+\nu_\mu$
with the $\mu^+$ treated as a $\pi^+$, and 
$K^+\rightarrow\pi^+\pi^0\pi^0$).
The insert in Fig. \ref{fig:peems2} 
exhibits a $M_{\pi ee}$ mass resolution of $\sigma =5.7$ MeV,
in good agreement with the Monte Carlo simulation. The calculated
$M_{ee}$ resolution is $\sigma =4.8$ MeV nearly independent
of $M_{ee}$.

The acceptance for \pee\ and \dal\ events was
determined to be 0.73 \% and 0.85 \%, respectively,
with a Monte Carlo simulation, which included 
the geometry of beam line and spectrometer, and
the separately measured efficiencies and responses of scintillators,
proportional chambers, \v{C}erenkov and shower counters. The influence
of the efficiencies on the acceptance nearly cancels 
in the normalisation, since
\pee\ and \dal\ events contain the same final states, and their
different spatial distributions produce only minor differences, 
and are taken into account.
Figures \ref{fig:dalitz} and \ref{fig:peems2} are examples of the many
control plots which attest the excellent quality of the simulation.
For \dal\ events the matrix elements of \cite{mikaelian} were
used for $\pi^0\to e^+e^-\gamma$, and the theoretical input to the 
\pee\ simulation is discussed below. 
Figure \ref{fig:dalitz} also demonstrates that the major difference between
\pee\ and \dal\ events, the high-mass trigger, 
is correctly accounted for, in magnitude as well
as in shape. 

\begin{figure}[htb]
\epsfig{figure=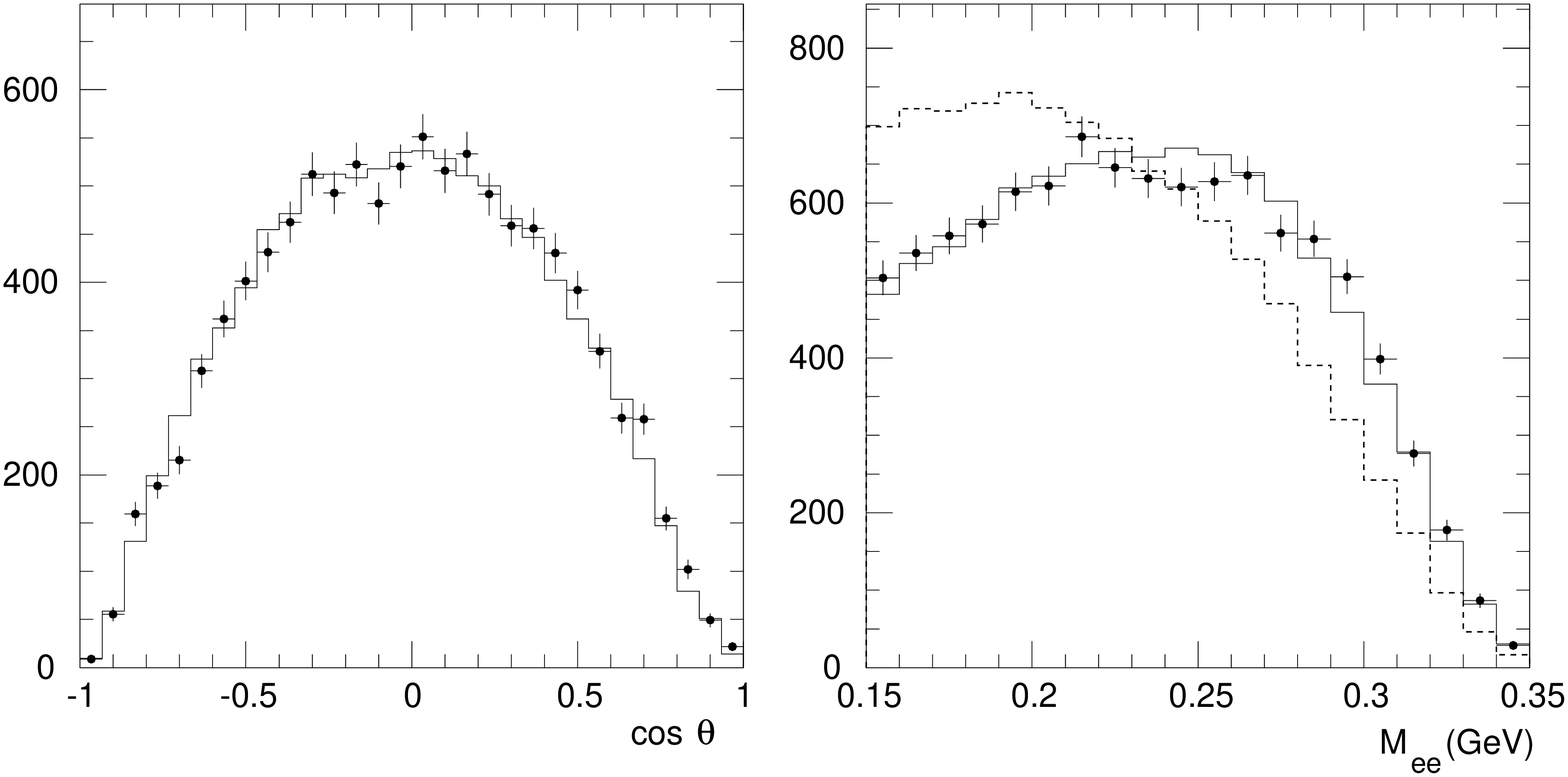,width=85mm,height=65mm}
\centering
\caption[]{Angular (left) and invariant mass
(right) distributions for \pee\ events (data points)
compared to Monte Carlo simulations (histogram)
assuming a pure vector interaction. A linear form factor
parametrisation with $\delta=2.14$ is used. The dashed histogram
(right) corresponds to a constant form factor ($\delta=0$).}
\label{fig:pieetheta}
\end{figure}

The essential distributions necessary for the interpretation
of our data are shown in Fig. \ref{fig:pieetheta}. Since the decay 
is supposed to proceed through one photon exchange, 
{\em i.e.} by a vector interaction (V) 
with a decay amplitude\cite{dambrosio} 
\[ \frac{\alpha G_F}{4\pi}f_V(z)P^\mu\overline{u}_e\gamma_\mu u_e\ , \]
one expects an angular distribution proportional to $\sin^2\theta$, 
where $\theta$ is the angle between the positron and pion 
momentum vectors in the center of mass of the $e^+e^-$ pair. 
The presence of other decay mechanisms, 
however, may produce small admixtures of scalar
(S) or tensor (T) terms \cite{macdowell}. The corresponding decay amplitudes 
\[ G_Fm_Kf_S\overline{u}_eu_e\;\;\;{\rm or}\;\;\; 
G_Ff_T\frac{P^\mu q^\nu}{m_K}
\overline{u}_e\sigma_{\mu\nu}u_e\;\ , \]
lead to either a constant (S) angular distribution or one
proportional to $\cos^2\theta$ (T). 
Here $G_F$ is the Fermi constant, $P=p_K+p_\pi$,
$q=p_K-p_\pi$, and the form factors $f_{V,S,T}$ are 
dimensionless functions of 
$z=M_{ee}^2/m_K^2$.  
Fitting a two-dimensional distribution
whose projections are shown in Fig. \ref{fig:pieetheta}, 
we find good agreement with a vector interaction. At 90 \%
confidence level at most 2 \% of the branching ratio could result
from either scalar or tensor interaction corresponding
to constraints $|f_S|<6.6\cdot 10^{-5}$ and $|f_T|<3.7\cdot 10^{-4}$.

Neglecting terms proportional to $m_e^2/M_{ee}^2$
the mass distribution for a vector interaction 
can be described by 
\begin{equation}
\frac{d\Gamma}{dz}=\frac{G_F^2\alpha^2m_K^5}{3(4\pi)^5}
\eta^{\frac{3}{2}}(1,z,m_\pi^2/m_K^2)|f_V(z)|^2\ ,\label{eq:decay}
\end{equation}
\noindent where $\eta(a,b,c)=a^2+b^2+c^2-2ab-2ac-2bc$.
The form factor $f_V(z)$ can be determined by fitting the 
observed spectrum in the experimentally accessible range $0.1<z<0.51$.
We have used two different parametrisations of
the form factor, one model independent (Eq.
\ref{eq:lambda}) and the other
derived from ChPT \cite{dambrosio} (Eq. \ref{eq:chpt}):
\begin{eqnarray}
f_V(z) & = & f_0(1+\delta z+\delta^\prime z^2)\label{eq:lambda}\ , \\
f_V(z) & = & a_++b_+z+w^{\pi\pi}(z)\ .\label{eq:chpt} 
\end{eqnarray}
$f_0,\delta,\delta^\prime,a_+, b_+$ are free parameters \cite{notation},
and $w^{\pi\pi}$ is the contribution from a pion loop graph
given in 
\cite{dambrosio}.

\begin{figure}[htb]
\epsfig{figure=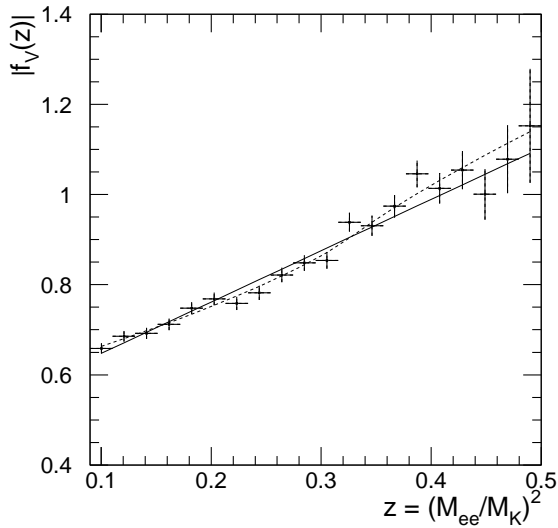,width=80mm}\centering
\caption[Form factor fits]{The measured form factor $|f_V(z)|$ versus
$z$. The solid line shows the linear fit 
(Eq.~\ref{eq:lambda} with $\delta^\prime=0$), 
the dashed line the next-to-leading-order ChPT fit 
(Eq.~\ref{eq:chpt}).}
\label{fig:fits}
\end{figure}

Figure \ref{fig:fits} displays the form factor, which
is extracted from the ratio of mass distributions for
the measured events to events simulated with a constant form factor.
The results of our fit to the linear
and ChPT ansatz are superimposed. Figure \ref{fig:pieetheta} shows the
spectrum itself. The parameters and branching ratios
are given in Table \ref{tab:fits}. The different contributions
to the systematic uncertainty of our results are listed in
in Table \ref{syserr}. Radiative corrections have been included
in the simulation following 
\cite{lautrup}. This increased the
branching ratio by 5.5 \% and the linear slope by 4 \%.

\begin{table}[hb]
\begin{tabular}{lcclc} 
\multicolumn{3}{c}{\hspace*{10mm}Eq. \ref{eq:lambda}}
& \multicolumn{2}{c}{Eq. \ref{eq:chpt}} \\\hline 
\multicolumn{1}{r}{$f_0$} & $0.533\pm 0.012$ 
& $0.591\pm 0.027$ & $a_+$ & $-0.587\pm 0.010$\\
\multicolumn{1}{r}{$\delta$ }& $2.14\pm 0.13$ 
& $0.97\pm 0.44$ & $b_+$ & $-0.655\pm 0.044$\\ 
\multicolumn{1}{r}{$\delta^\prime$} & 0 & $1.99\pm 0.67$ & & \\
$\chi^2/n_{\rm dof}$ & 22.9/18 & 16.6/17 & & 13.3/18  \\\hline
$BR$ (total) & $2.884\pm  0.037$ & $2.991\pm  0.058$ && $2.988 \pm 0.040$ \\
$BR$ (meas.)& \multicolumn{4}{c}{$2.015\pm 0.020$}\\
\end{tabular}
\caption[Fits to the present data.]{Summary of results
for the fits to the measured form factor.
The branching ratios ($BR$) are given in units of $(10^{-7})$.
$BR$ (meas.) corresponds to $M_{ee}>0.15\;{\rm GeV}$ 
and is within the statistical error
independent of the parametrisation of the form factor.
\label{tab:fits}}\vspace*{3mm}
\begin{tabular}{lcc}
Source & $\sigma_{BR}/BR$ (\%) & $\sigma_\delta/\delta$ (\%)
\\\hline
Normalisation to 
$\pi^0\rightarrow e^+e^-\gamma$ & 2.8 & \\
Radiative corrections & 1.4 & 0.5 \\
Reconstruction & 3.0 & 6.5 \\
Background subtraction & 1.5 & 2.3 \\
\hline
Total & 4.5 & 6.9 \\
\end{tabular}
\caption[Systematic errors]{Summary of systematic errors.
The reconstruction and background subtraction error
result from adding several smaller contributions linearly.
The total error is the sum of the four groups in quadrature.
\label{syserr}}
\end{table}

The inclusion of the quadratic term in the model independent
ansatz improves the quality of the fit. 
The values of $\delta$ and $\delta^\prime$ are however
strongly correlated. Our result $\delta=2.14\pm 0.13\pm 0.15$
is in fair agreement with the less precise results of experiment
E777~\cite{alliegro,notation} $\delta=1.31\pm 0.44$.
The large value of $\delta$ is in contradiction with 
the meson dominance class of models \cite{lichard,bergstroem},
in which the form factor slope is given in first
approximation as $\delta=(m_K/m_\rho)^2$.

The ChPT parametrisation of Eq. \ref{eq:chpt} includes
the contribution from the pion loop graph. This term has a strong
$z$ dependence, which has not been calculated explicitly in most
earlier models. In fact, in the first, lowest non-trivial order 
ChPT calculation (${\cal O}(p^4)$\cite{ecker}) 
this term was assumed to be the only significant source of $z$
dependence of the form factor ({\em i.e.} $b_+=0$). Our data show
that this is a poor approximation.
Significant $z$ dependence from other terms is expected
in next-to-leading-order ChPT~\cite{donoghue,dambrosio}.
The amplitude linear in $z$ (Eq. \ref{eq:chpt})
is thought to represent all
contributions other than the pion loop term, which
ChPT cannot calculate \cite{dambrosio}.
The substantial reduction 
in the value of $\chi^2$ which is observed when the
pion loop term is included provides direct
experimental evidence for its small but important contribution.

Summarizing the results of our analysis we conclude: 
(i) the experimental data are consistent with a 
vector model for the interaction, 
(ii) the slope of the form factor is significantly larger
than meson dominance or leading order ChPT models predict, 
and (iii) although a linear approximation
of the form factor is reasonable, 
our data indicate a nonlinearity of the form factor
which is fit well by the ChPT loop term. 
Our final result for the total branching ratio is
$(2.94\pm 0.05\,(stat.)\pm 0.13\,(syst.)
\pm 0.05\,(model))\times 10^{-7}$.
Here we include the model dependence
of the extrapolation into the low mass region not covered by our detector
by taking the average of the extreme values in Table~\ref{tab:fits}.

We gratefully acknowledge the contributions to the success of
this experiment by Dave Phillips,  
the staff and management of the AGS at the Brookhaven National
Laboratory, and the technical staffs of the participating institutions.
This work was supported in part by the U. S. Department of Energy, 
the National Science Foundations of the USA, Russia and Switzerland, and
the Research Corporation.

\end{document}